\documentclass[journal,onecolumn]{IEEEtran}

\usepackage{amsmath, amsfonts, amssymb, float, fancyhdr}
\usepackage[figuresright]{rotating}
\usepackage{authblk, graphicx, indentfirst, lastpage, lipsum}
\usepackage{pifont}

\makeatletter
\renewcommand{\@biblabel}[1]{[#1]\hfill}
\makeatother
\usepackage{subfig, caption, epstopdf}
\usepackage[left=3cm, right=2.5cm, top=2.5cm, bottom=2.5cm, includehead, includefoot]{geometry}
\usepackage[justification=centering]{caption}
\usepackage{titlesec}
\usepackage{xcolor}
\titleformat{\section}
{\normalfont\normalsize\bfseries\uppercase}{\thesection}{1.7em}{}
\titleformat{\subsection}
{\normalfont\normalsize\bfseries}{\thesubsection}{.95em}{}
\titleformat{\subsubsection}
{\bfseries}{\thesubsubsection}{.2em}{}
\titlespacing*{\section}{0cm}{0.7cm}{0cm}
\usepackage{enumitem}
\usepackage[numbers,compress]{natbib}
\usepackage{multirow}

\author[1]{\bfseries Muhammad Wildan Oktavian}
\author[2]{\bfseries Novanto Yudistira}
\author[3]{\bfseries Achmad Ridok}

\affil[1,2,3]{Informatics Engineering, Faculty of Computer Science, Brawijaya University, Indonesia}

\title{Classification of Alzheimer's Disease Using the Convolutional Neural Network (CNN) with Transfer Learning and Weighted Loss}

\begin{document}
\setcounter{page}{1}

\setlength{\parindent}{1.27cm}

\pagestyle{fancy}
\fancyhfoffset{0cm}


\maketitle





\begin{abstract}
Alzheimer's disease is a progressive neurodegenerative disorder that gradually deprives the patient of cognitive function and can end in death. With the advancement of technology today, it is possible to detect Alzheimer's disease through Magnetic Resonance Imaging (MRI) scans. So that MRI is the technique most often used for the diagnosis and analysis of the progress of Alzheimer's disease. With this technology, image recognition in the early diagnosis of Alzheimer's disease can be achieved automatically using machine learning. Although machine learning has many advantages, currently the use of deep learning is more widely applied because it has stronger learning capabilities and is more suitable for solving image recognition problems. However, there are still several challenges that must be faced to implement deep learning, such as the need for large datasets, requiring large computing resources, and requiring careful parameter setting to prevent overfitting or underfitting. In responding to the challenge of classifying Alzheimer's disease using deep learning, this study propose the Convolutional Neural Network (CNN) method with the Residual Network 18 Layer (ResNet-18) architecture. To overcome the need for a large and balanced dataset, transfer learning from ImageNet is used and weighting the loss function values so that each class has the same weight. And also in this study conducted an experiment by changing the network activation function to a mish activation function to increase accuracy. From the results of the tests that have been carried out, the accuracy of the model is 88.3\% using transfer learning, weighted loss and the mish activation function. This accuracy value increases from the baseline model which only gets an accuracy of 69.1\%.
\end{abstract}



\section{Introduction}
\label{}
As people age, human health will inevitably decline, making them susceptible to disease. One part of the body that is the main target of the effects of aging is the brain. The brain will experience changes in intellectual function, such as difficulty remembering and being slow to take a decision or action [1]. One of the diseases caused by this brain disorder is Alzheimer. Alzheimer's disease is a progressive neurodegenerative disorder that gradually robs the patient's cognitive function and can end in death [2]. This disease is generally a cause of dementia in the elderly. The majority of people with Alzheimer will experience several disorders such as memory disorders, changes in character or personality, moods, and difficulty interacting with other people [3]. Someone diagnosed with Alzheimer will gradually experience several disorders over three to nine years [4].

With the advancement of technology today, it is possible to detect Alzheimer's disease through Magnetic Resonance Imaging (MRI) scans [5]. Typically, MRI is the technique most often used for the diagnosis and analysis of the progress of Alzheimer's disease [6]. With this technology, early diagnosis of Alzheimer's disease can be achieved automatically using machine learning. In some cases, machine learning can predict Alzheimer's disease better than medical personnel, making this field important for computer-based diagnostic research [7]. Although machine learning has many advantages, it is unsuitable for image recognition. Currently, the field of image recognition uses deep learning, which has stronger learning abilities and is more suitable for solving image recognition problems [8]. Many deep learning methods outperform machine learning methods in image recognition, such as Convolutional Neural Network (CNN) and sparse autoencoder [9]. Even though deep learning has robust performance, there are still some challenges. Firstly, it requires a large amount of training data which can be a problem because medical data is expensive and protected from cross-institutional use for ethical reasons. Training deep learning networks with large amounts of image data requires large computing resources. Deep learning networks require careful parameter settings because non-optimal parameter settings can result in overfitting or underfitting, which leads to poor network performance [10].

In detecting the results of MRI images using deep learning, many approaches have been taken. As in the previous study on the classification of Alzheimer's disease by Aly Valliani and Ameet Soni in 2017 [11] entitled "Deep Residual Nets for Improved Alzheimer's Diagnosis" which found that the use of transfer learning and augmentation can improve classification accuracy on the CNN Deep Residual Nets architecture. Research on a similar topic was also conducted by Aderghal et al. [12] under the title "Classification of Alzheimer's Disease on Imaging Modalities with Deep CNNs using Cross-Modal Transfer Learning" using transfer learning. It reduces overfitting, improves model performance, and increases prediction accuracy. Then there is also a study on class imbalance, such as Songqing Yue in 2017 [13] entitled "Imbalanced Malware Images Classification: a CNN based Approach" which uses a weighted loss on the CNN architecture. It reduces the effect of class imbalance on the dataset used to increase accuracy. From the problem exposure and previous research, the we proposed Convolutional Neural Network (CNN) method using Residual Neural Network (ResNet) architecture with Weighted Loss and Transfer Learning to classify Alzheimer's disease from MRI data into three classes.

\section{Related Work}
\label{}
Lucas R. Trambaiolli, Ana C. Lorena, Francisco J. Fraga, Paulo A.M. Kanda, Renato Anghinah, and Ricardo Nitrini 2011 [14] researched the classification of Alzheimer's Disease entitled "Improving Alzheimer's Disease Diagnosis with Machine Learning Techniques". Researchers used the Support Vector Machine (SVM) machine learning method to differentiate between Alzheimer's patients and Controlled patients. From the results of the analysis that has been done, the accuracy is 79.9\% and the sensitivity is 83.2\%. This study suggests using more data and reconsidering the parameters of the SVM classifier to improve the resulting performance.

Aly Valliani and Ameet Soni 2017 [11] researched the classification of Alzheimer's disease entitled "Deep Residual Nets for Improved Alzheimer's Diagnosis." The researcher focuses on overcoming the problem of the amount of data to train the CNN model. Therefore, the researcher uses the Deep Residual Nets (ResNet) architecture, which was previously trained with the ImageNet dataset, which has large-scale image data. The results of this study indicate that the use of transfer learning from ImageNet and augmentation can improve classification accuracy. The model gets a test accuracy of 81.3\% for binary classification and 56.8\% for multiclass or 3-way classification. This result is better than the model that does not use transfer learning and augmentation, with an accuracy of 78.8\% for binary classification and 56.1\% for 3-way classification.

Aderghal et. al in 2018 [12] researched the classification of Alzheimer's disease entitled "Classification of Alzheimer's Disease on Imaging Modalities with Deep CNNs using Cross-Modal Transfer Learning". Researchers stated that public data on Alzheimer's disease is not much, so it will result in the phenomenon of overfitting when trained. Therefore, researchers propose using transfer learning from larger datasets to improve classification accuracy. The results of this study indicate that the use of transfer learning on the CNN model can improve the model's performance, reduce overfitting, and increase classification accuracy. Furthermore, the classification can be leveraged to perform multi instance learning on Alzheimer's disease dataset to localize benign and malignant part of the brain (Kavitha et. al 2019 [26]).

Songqing Yue 2017 [13] conducted a study entitled "Imbalanced Malware Images Classification: a CNN-based Approach". According to this study, CNN classification performance will decrease when the dataset has an unbalanced number of classes or Class Balance. To overcome this problem, the loss value weighting in the last CNN layer is used. With this weighting, the misclassification of the minority class will be strengthened, and the majority class will be reduced so there can be a balance. Yudistira et. al 2020 [27] conducted research on organelles localization prediction using U-Net under imbalanced dataset. It employs weighting term on loss function to detect minority organelles.  

Krit Sriporn, Cheng-Fa Tsai, Chia-En Tsai, and Paohsi Wang 2020 [15] conducted a study entitled "Analyzing Lung Disease Using Highly Effective Deep Learning Techniques". This study used the mish activation function to replace the ReLU activation function in several well-known architectures such as MobileNet, Densenet-121, and Resnet-50. This mish activation function is the current state-of-the-art activation function. From the analysis results, it was found that the mish activation function can increase the classification accuracy to 98.88\% from the model that does not use mish activation or the baseline model with an accuracy of 97.25\%.

\section{Research Method and Materials}
\label{}
\subsection{Deep Residual Network}
According to Kaiming He [16], in general, the deeper the structure of the neural network, the more difficult it will be to learn. However, Residual Neural Network (ResNet) provides a residual learning framework to simplify the training process even though it uses a deep network structure. ResNet explicitly reformulates the network layer into residual learning functions that lead to the input layer rather than learning unreferenced functions. As the deeper layers of the network begin to converge, degradation problems arise, where as the network depth increases, the accuracy saturates and then decreases rapidly. The degradation is not caused by vanishing gradients, overfitting and adding more layers to the model leads to a higher training error. The architectural type of ResNet is distinguished by the number of layers on the network. Among the architectures used in testing for the ILSVRC competition are ResNet-18, ResNet-34, ResNet-50, ResNet-101, and ResNet-152.

\subsection{Weighted Cross Entropy Loss}
Cross Entropy is a measure of the field of information theory, which builds on entropy and usually calculates the difference between two probability distributions. Cross Entropy is also associated with and is often misunderstood as logistic loss, which is commonly referred to as log loss. Although the two measures come from different sources, when used as a loss function for a classification model, they both calculate the same quantity and can be used interchangeably [17]. the formula used to calculate the cross entropy loss can be seen in Equation 1 below.

\begingroup
\Large
\begin{equation}
-\sum_{c=1}^{M} y_{o,c} \log(P_{o,c})
\end{equation}
\vspace{.005em}
\endgroup


where \(M\) is the number of existing classes, \(y_{o,c}\) is a binary indicator (0 or 1) if the class c label is the correct classification for the sample o, and \(p_(o,c)\) is the predicted probability sample o from class c. According to Naceur et. al.[18], to overcome the class imbalance in the dataset, the formula for the loss function will be added with a weight based on the number of samples from each class. The formula used to calculate the weighted cross entropy loss can be seen in Equation 2 and Equation 3 below.

\begingroup
\Large
\begin{equation}
-\sum_{c=1}^{M} W_{o,c} y_{o,c} \log(P_{o,c})
\end{equation}
\vspace{.005em}
\endgroup

\begingroup
\Large
\begin{equation}
W_{o,c} = 1 - \frac{x_{c}}{N}
\end{equation}
\vspace{.005em}
\endgroup

where \(W_(o,c)\) is the specific weight for each class c, \(x_{c}\) is the number of samples in class c, and \(N\) is the total of all samples from all classes

\subsection{Proposed Method}

\begin{figure}[H]
    \centering
    \includegraphics[scale=0.35]{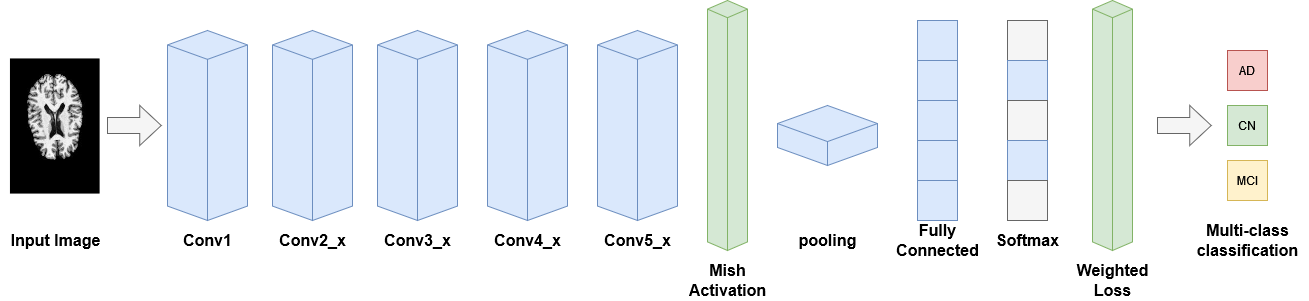}
    \vspace{.7em}
    \caption{An Overview of our approach}%
\end{figure}
As shown in Figure 1 above, we use the main part of the ResNet architecture for the training process with modifications. We modified the activation function used in the last residual block before the pooling process. We use the Mish Activation Function to replace the ReLU activation function which is the default activation function of the ResNet architecture. Mish is a non-monotonic activation function that is smooth, continuous, and self regularized. Inspired by the Swish Activation Function, Mish uses the Self-Gating property where the non-modulated input is multiplied by the output of the non-linear function of the input[19]. The formula for the Mish Activation function can be seen in Equation 4 below::

\begingroup
\Large
\begin{equation}
f(x) = x tanh(softplus(x)) = x tanh (ln(1 + e^x))
\end{equation}
\vspace{.005em}
\endgroup

Then we calculate the loss value using the weighted cross entropy loss. The formula for calculating this loss value can be seen in Equation 2. This weighted cross entropy is used because the dataset we use has an imbalance in the number of samples. This sample consists of AD, MCI, and CN classes. Class AD has the least number of these three classes. Then the number of samples from these three classes will be used to calculate the weights in Equation 3.

\subsection{Data Acquisition and Preprocessing}
Data used in the preparation of this article were obtained from the Alzheimer’s Disease Neuroimaging Initiative (ADNI) database (adni.loni.usc.edu) [20]. The ADNI was launched in 2003 as a public-private partnership, led by Principal Investigator Michael W. Weiner, MD. The primary goal of ADNI has been to test whether serial magnetic resonance imaging (MRI), positron emission tomography (PET), other biological markers, and clinical and neuropsychological assessment can be combined to measure the progression of mild cognitive impairment (MCI) and early Alzheimer’s disease (AD). These imaging data are acquired from 306 ADNI
participants including 133 mild cognitive impairment (MCI), 58 Alzheimer Disease (AD), and 115 Normal Control (CN). for each image has a size of 256 x 256 x 256 which will be split into 256 slices.

\begin{figure}[H]
    \centering
    \subfloat[\centering ]{{\includegraphics[scale=0.6]{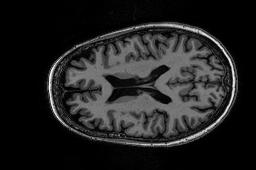} }}%
    \qquad
    \subfloat[\centering ]{{\includegraphics[scale=0.6]{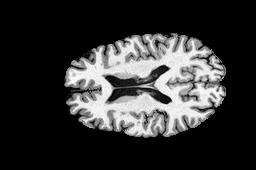} }}%
    \vspace{.7em}
    \caption{Comparison between unpreprocessed and preprocessed data sample: (a) Unpreprocessed Data, (b) Preprocessed Data}%
\end{figure}

\begin{figure}[H]
    \centering
    \subfloat[\centering ]{{\includegraphics[scale=0.6]{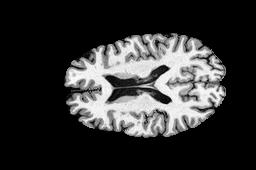} }}%
    \qquad
    \subfloat[\centering ]{{\includegraphics[scale=0.6]{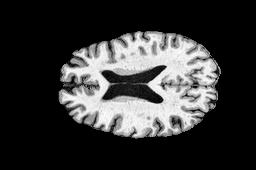} }}%
    \qquad
    \subfloat[\centering ]{{\includegraphics[scale=0.6]{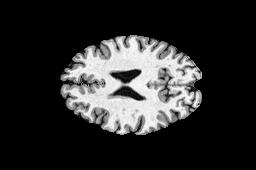} }}%
    \vspace{.7em}
    \caption{Preprocessed dataset samples : (a) AD Sample, (b) CN Sample, (c) MCI Sample}%
\end{figure}

The dataset file obtained is a dataset in the form of 3-dimensional images in nifti format. Before it can be used for training, it needs to be preprocessed. Image data will be segmented first to remove the skull and other parts of the head that are not brain. This segmentation aims so that the model only focuses on the brain. This segmentation is done using the DeepBrain library [21].Then after segmentation, slicing or cutting of 3-dimensional images into 2 dimensions will be carried out using the med2image library [22] to be used in network training. The difference in the image data before and after preprocessing can be seen in figure 2 and the results of preprocessing for each class can be seen in figure 3. From Figures 2 and 3 it is also seen that the preprocessed image data has brighter and sharper colors. Then the preprocessed dataset will then be divided into 3 parts, namely training data, validation data and test data. The process of dividing this dataset using K-Fold with K totaling 5. First, the dataset will be divided into 2 parts with a composition of 20\% test data and 80\% training data. Then the 80\% training data will be further divided into 80\% training data and 20\% validation data.



\section{Result and Discussion}
\label{}

\subsection{ResNet Architecture Comparison}
The capacity of the neural network to learn is determined by the architecture used. By using the appropriate architecture, the network model can learn data patterns better. The scenario for testing the model architecture can be seen in the Table 1 below.

\begin{table}[H]
    \centering
    \fontsize{8pt}{10pt}
    \selectfont
    \caption{ResNet Architecture Comparison}
    \begin{tabular}{|c|c|c|c|c|c|c|c|c|c|c|}
        \hline
        \multirow{3}{*}{K-Fold} & \multicolumn{10}{c|}{Accuracy} \\
        \cline{2-11}  
        & \multicolumn{2}{|c}{ResNet-18} & \multicolumn{2}{|c|}{ResNet-34} & \multicolumn{2}{c|}{ResNet-50} & \multicolumn{2}{c|}{ResNet-101} & \multicolumn{2}{c|}{ResNet-152} \\
        \cline{2-11} 
        & Validation & Test & Validation & Test & Validation & Test & Validation & Test & Validation & Test\\
        \hline
        1 & 0,523 & 0,531 & 0,461 & 0,879 & 0,491 & 0,823 & 0,448 & 0,487 & 0,675 & 0,473 \\
        2 & 0,549 & 0,897 & 0,497 & 0,88 & 0,472 & 0,504 & 0,494 & 0,849 & 0,582 & 0,381\\
        3 & 0,578 & 0,655 & 0,601 & 0,513 & 0,6 & 0,472 & 0,433 & 0,8 & 0,525 & 0,597\\
        4 & 0,523 & 0,835 & 0,575 & 0,669 & 0,61 & 0,459 & 0,49 & 0,836 & 0,404 & 0,7\\
        5 & 0,589 & 0,534 & 0,561 & 0,489 & 0,542 & 0,881 & 0,553 & 0,414 & 0,53 & 0,503\\
        \hline
        Mean & \textbf{0,553} & \textbf{0,691} & 0,539 & 0,686 & 0,543 & 0,628 & 0,484 & 0,677 & 0,543 & 0,531\\
        \hline
    \end{tabular}
\end{table}

From the results of the architectural tests carried out on the accuracy of the validation data, data testing, and the duration of the training, it shows that the deeper the network layer of the architecture used the accuracy tends to decrease, although not significantly. This is due to overfitting because the architecture used is increasingly complex so that the accuracy decreases and the training time becomes longer. The results of this test indicate that in this study the ResNet-18 architecture has better performance with the shortest training time.

\subsection{Optimizer Comparison}
\label{}
The optimization algorithm or optimizer was tested 3 times, namely using Stochastic Gradient Descents (SGD), Root Mean Square Propogation (RMSprop), and Adam Optimizer. In the SGD optimization, the Learning Rate parameter of 0.001 and the momentum of 0.9 , in the RMSprop optimization the Learning Rate parameter of 0.01, the alpha of 0.99, the epsilon of 1e-08 weight decay of 0, and the momentum of 0, and the Adam optimization the Learning parameter is used Rate is 0.001, beta from 0.9 to 0.999, epsilon is 1e-08, weight decay is 0.

\begin{table}[H]
    \centering
    \fontsize{8pt}{10pt}
    \selectfont
    \caption{Optimizer Comparison}
    \begin{tabular}{|c|c|c|c|c|c|c|}
        \hline
        \multirow{3}{*}{K-Fold} & \multicolumn{6}{c|}{Accuracy} \\
        \cline{2-7}  
        & \multicolumn{2}{|c}{SGD} & \multicolumn{2}{|c|}{RMSprop} & \multicolumn{2}{c|}{Adam} \\
        \cline{2-7} 
        & Validation & Test & Validation & Test & Validation & Test\\
        \hline
        1 & 0,523 & 0,531 & 0,519 & 0,769 & 0,604 & 0,444 \\
        2 & 0,549 & 0,897 & 0,645 & 0,438 & 0,646 & 0,437 \\
        3 & 0,578 & 0,655 & 0,596 & 0,446 & 0,600 & 0,463 \\
        4 & 0,523 & 0,835 & 0,485 & 0,879 & 0,530 & 0,503 \\
        5 & 0,589 & 0,534 & 0,644 & 0,445 & 0,651 & 0,456 \\
        \hline
        Mean & 0,553 & \textbf{0,691} & 0,578 & 0,595 & \textbf{0,606} & 0,460\\
        \hline
    \end{tabular}
\end{table}

From the tests that have been carried out in the Table 2 above, it shows that the SGD optimization algorithm gives the best results compared to other optimization algorithms on the test data. As for the validation data, the highest accuracy is obtained by the adam optimization algorithm. Because what is used as a reference in this research is test data, so SGD will be used as an optimization algorithm in the next test.

\subsection{Weighted Loss Function}
In datasets that have an unbalanced number of classes, it will affect the results of the classification carried out. For this reason, it is necessary to do class weighting on the network loss function so that the class that has fewer numbers will have a greater weight than the number of classes that have more. This test was carried out using the ResNet-18 architecture from the results of the previous architecture test. The scenario for testing this class weighting can be seen in the Table 3 below.

\begin{table}[H]
    \centering
    \fontsize{8pt}{10pt}
    \selectfont
    \caption{Weighted Loss Function}
    \begin{tabular}{|c|c|c|c|c|c|c|c|c|c|c|c|}
        \hline
        \multirow{2}{*}{Condition} & \multicolumn{2}{c|}{Accuracy} & \multicolumn{3}{c|}{Precision} & \multicolumn{3}{c|}{Recall} & \multicolumn{3}{c|}{F1-Score} \\
        \cline{2-12}  
        & Validation & Test & AD & CN & MCI & AD & CN & MCI & AD & CN & MCI\\
        \hline
        Baseline & 0,553 & 0,691 & 0,550 & 0,490 & \textbf{0,916} & 0,761 &	\textbf{0,857} & 0,654 & 0,598 & 0,589 & 0,707\\
        \hline
        Weighted Loss & \textbf{0,574} & \textbf{0,799} & \textbf{0,645} & \textbf{0,771} & 0,893 & \textbf{0,899} & 0,853 & \textbf{0,775} & \textbf{0,672} & \textbf{0,795} & \textbf{0,823} \\
        \hline
    \end{tabular}
\end{table}

From the results of the tests carried out in the Table 3 above, it shows that applying class weighting has better results. The model can recognize AD and CN classes better which is indicated by the increasing precision value obtained. From the F1 Score obtained, it also shows that each class gets an increase in value. This shows that class weighting can improve the performance of the model in classifying.

\subsection{Transfer Learning}
Network models that have been trained previously using larger datasets (pretrained) can help other network models in helping to get better accuracy. The weights from the results of the pretrained model will be used to train other models with the new dataset. In this test, the ResNet-18 architecture is used from the results of the previous architecture test and the transfer learning weights obtained from ImageNet. The transfer learning test scenario can be seen in the Table 4 below.

\begin{table}[H]
    \centering
    \fontsize{8pt}{10pt}
    \selectfont
    \caption{Transfer Learning}
    \begin{tabular}{|c|c|c|c|c|c|c|c|c|c|c|c|}
        \hline
        \multirow{2}{*}{Condition} & \multicolumn{2}{c|}{Accuracy} & \multicolumn{3}{c|}{Precision} & \multicolumn{3}{c|}{Recall} & \multicolumn{3}{c|}{F1-Score} \\
        \cline{2-12}  
        & Validation & Test & AD & CN & MCI & AD & CN & MCI & AD & CN & MCI\\
        \hline
        Baseline & \textbf{0,553} & 0,691 & 0,550 & 0,490 & \textbf{0,916} & 0,761 &	0,857 & 0,654 & 0,598 & 0,589 & 0,707\\
        \hline
        Transfer Learning & 0,552 & \textbf{0,895} & \textbf{0,909} & \textbf{0,944} & 0,849  & \textbf{0,845} & \textbf{0,884} & \textbf{0,933} & \textbf{0,874} & \textbf{0,912} & \textbf{0,889} \\
        \hline
    \end{tabular}
\end{table}

From the results of the tests carried out in the Table 4 above, it shows that the use of transfer learning significantly increases the accuracy, precision, recall and f1 score obtained.

\subsection{Mish Activation}
In this section, testing is carried out in the form of changing the activation function which previously used the ReLu function to be converted into the Mish function. The network layer whose activation function is changed is divided into several conditions, namely the first is the last convolution layer before entering the fully connected layer, and the second changes all activation functions on the network into Mish functions. These two conditions will be compared with the initial architecture that uses the ReLu activation function as the standard condition. The scenario for testing the effect of using the Mish activation function can be seen in the Tables 5 below.

\begin{table}[H]
    \centering
    \fontsize{8pt}{10pt}
    \selectfont
    \caption{Mish Activation}
    \begin{tabular}{|c|c|c|c|c|c|c|c|c|c|c|c|}
        \hline
        \multirow{2}{*}{Condition} & \multicolumn{2}{c|}{Accuracy} & \multicolumn{3}{c|}{Precision} & \multicolumn{3}{c|}{Recall} & \multicolumn{3}{c|}{F1-Score} \\
        \cline{2-12}  
        & Validation & Test & AD & CN & MCI & AD & CN & MCI & AD & CN & MCI\\
        \hline
        Baseline & \textbf{0,567} & \textbf{0,885} & 0,871 & 0,918 & \textbf{0,855} & 0,830  & \textbf{0,904} & 0,911 & 0,849 & \textbf{0,909} & 0,879\\
        \hline
        Mish (Last Layer) & 0,531 & 0,883 & \textbf{0,901} & \textbf{0,930} & 0,846 & 0,801 & 0,890 & \textbf{0,920} & 0,844 & \textbf{0,909} & \textbf{0,882} \\
        \hline
        Mish (All Layer) & 0,547 & 0,879 & 0,882 & 0,919 & 0,845 & \textbf{0,831} & 0,875 & 0,919 & \textbf{0,856} & 0,893 & 0,878\\
        \hline
    \end{tabular}
\end{table}

From the results of the tests carried out in the Table 5 above, it shows that the model that uses a combination of weighted loss and transfer learning with the mish activation function is not better than the baseline model. The use of mish function and weighted loss does not increase the accuracy of the model that already uses transfer learning. Although the model with the mish activation function in the last layer of the network has slightly lower accuracy, this model can increase the precision for the AD class by 3\% and 1.2\% for the CN class.

\begin{figure}[H]
    \centering
    \subfloat[\centering ]{{\includegraphics[scale=0.15]{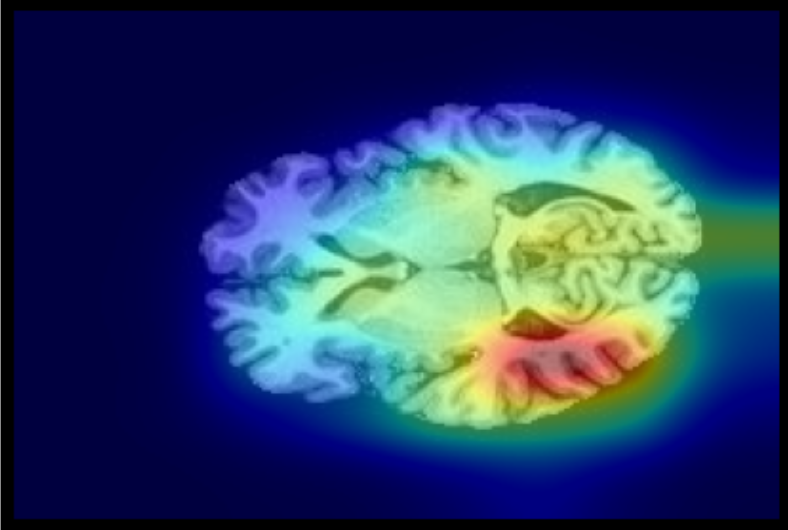} }}%
    \qquad
    \subfloat[\centering ]{{\includegraphics[scale=0.15]{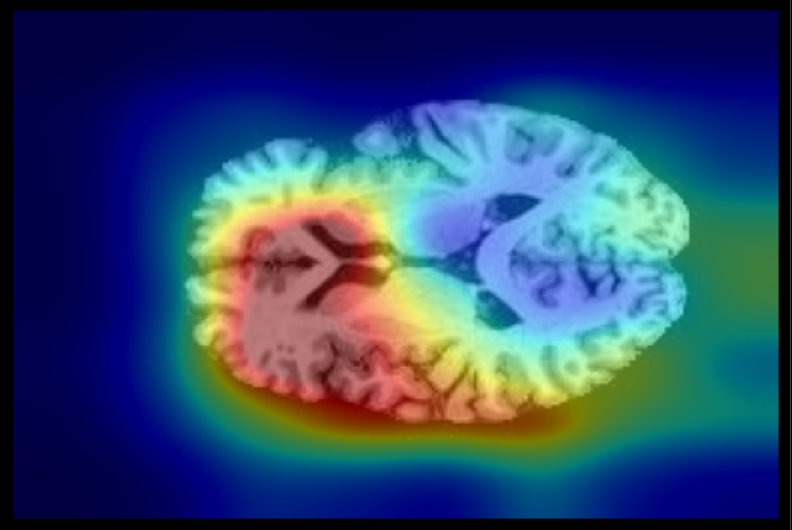} }}%
    \qquad
    \subfloat[\centering ]{{\includegraphics[scale=0.15]{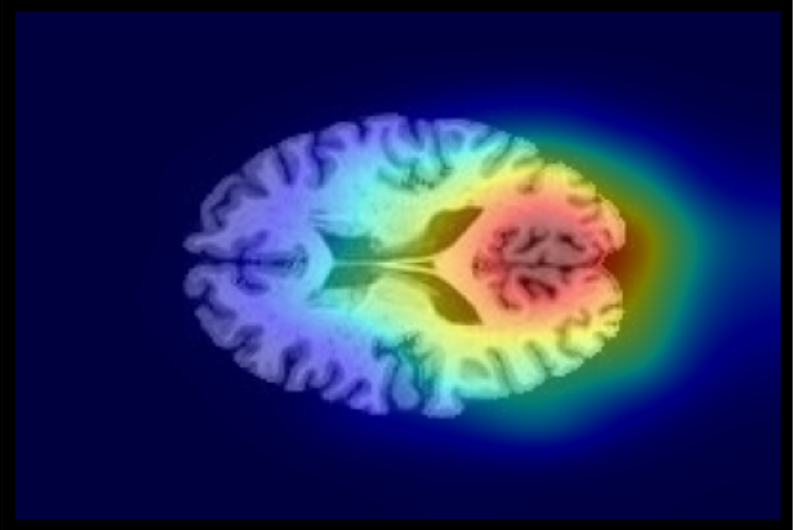} }}%
    \vspace{.7em}
    \caption{Model visualization with Grad-CAM : (a) AD Sample, (b) CN Sample, (c) MCI Sample}%
\end{figure}

As a visual explanation in Figure 4 above, Grad-CAM is used to check the classification results by the model. Gradient-weighted Class Activation Mapping (Grad-CAM), uses the gradients of any target concept, flowing into the final convolutional layer to produce a coarse localization map highlighting the important regions in the image for predicting the concept [23]. In the figure above, the model is able to focus on three brain areas in classifying Alzheimer's disease. Generally, Grad-CAM is used to distinguish several different objects, but in this case study the focus area of classification in brain images shows the same location in each class because the characteristics of each class can be in the same part of the brain. Of the many cases of Alzheimer's, the three areas in the figure above are the locations of the most common symptoms of Alzheimer's disease which is hippocampus, ventricles, and cortex [24], so it can be ensured that the model can recognize and classify Alzheimer's disease.

\subsection{Comparison with previous study}
\begin{table}[H]
    \centering
    \fontsize{8pt}{10pt}
    \selectfont
    \caption{Comparison with previous study}
    \begin{tabular}{|c|c|c|}
        \hline
        Author & Architecture & Accuracy\\
        \hline
        Trambaiolli et. al., 2011 & Support Vector Machine (SVM) & 79.9\% (multiclass)\\
        \hline
        Valliani, 2017 & ResNet-18 + Pretrain + augmentation & 56.8\% (multiclass) and 81.3\% (biner)\\
        \hline
        \multirow{1}{*}{} & \multirow{1}{*}{} & 85.07\% (VGG-16),\\
        Acharya et. al, 2021 & VGG-16, ResNet-50, Modified AlexNet & 75.25\% (ResNet-50), \\
         &  & 95.70\% (Alexnet)\\
         &  & (multiclass)\\
        \hline
        Proposed method & ResNet-18 + Weighted Loss + Transfer Learning + Mish Activation & 88,3\% (multiclass)\\
        \hline
    \end{tabular}
\end{table}

As shown in Table 6 above, research conducted by Trambaiolli et. al.[14] used as a basis for comparing experiments using traditional machine learning and deep learning. The proposed research has succeeded in providing a higher test accuracy, however, the data used is not the same. Then the research by Acharya et. al.[25], the proposed research can outperform 2 out of 3 architectures, namely VGG-16 and ResNet-50. The AlexNet architecture gets higher accuracy by making modifications such as using only 2 of the five convolution layers and the Adam optimizer. This study also uses a different dataset, namely, using the dataset contained in the Kaggle repository. Then in Valliani's research [11], the dataset and model settings used are the same as the proposed research, so the use of weighted loss and mish activation is proven to improve model performance.






\section{Conclusion}
\label{}
In this study, we successfully classified the AD, CN and MCI data with 88.3\% Accuracy and precision of AD and CN 90.1\% and 93\% respectively. We have found that in this case, the deeper the number of layers of the ResNet network that is used does not increase the model's performance and even slightly lowers the accuracy obtained. This is because the amount of data used is not too much, namely a number of 10,794 images that have been extracted from 306 subjects. The use of each of the weighted loss, transfer learning and mish activation functions in the network model can improve the model's performance better, such as increasing the precision of AD and CN classes. However, when the transfer learning model is added with weighted loss and mish activation function, it does not show a significant increase in the accuracy obtained.

\section*{Acknowledgement}
\label{}
Data collection and sharing for this project was funded by the Alzheimer's Disease Neuroimaging Initiative (ADNI) (National Institutes of Health Grant U01 AG024904) and DOD ADNI (Department of Defense award number W81XWH-12-2-0012). ADNI is funded by the National Institute on Aging, the National Institute of Biomedical Imaging and Bioengineering, and through generous contributions from the following: AbbVie, Alzheimer’s Association; Alzheimer’s Drug Discovery Foundation; Araclon Biotech; BioClinica, Inc.; Biogen; Bristol-Myers Squibb Company; CereSpir, Inc.; Cogstate; Eisai Inc.; Elan Pharmaceuticals, Inc.; Eli Lilly and Company; EuroImmun; F. Hoffmann-La Roche Ltd and its affiliated company Genentech, Inc.; Fujirebio; GE Healthcare; IXICO Ltd.; Janssen Alzheimer Immunotherapy Research \& Development, LLC.; Johnson \& Johnson Pharmaceutical Research \& Development LLC.; Lumosity; Lundbeck; Merck \& Co., Inc.; Meso Scale Diagnostics, LLC.; NeuroRx Research; Neurotrack Technologies; Novartis Pharmaceuticals Corporation; Pfizer Inc.; Piramal Imaging; Servier; Takeda Pharmaceutical Company; and Transition Therapeutics. The Canadian Institutes of Health Research is providing funds to support ADNI clinical sites in Canada. Private sector contributions are facilitated by the Foundation for the National Institutes of Health (www.fnih.org). The grantee organization is the Northern California Institute for Research and Education, and the study is coordinated by the Alzheimer’s Therapeutic Research Institute at the University of Southern California. ADNI data are disseminated by the Laboratory for Neuro Imaging at the University of Southern California.





\bibliographystyle{IEEEtran}

\end{document}